\documentclass[twocolumn]{jpsj2}
\usepackage{graphicx}
\def\runtitle{%
Tunneling conductance
in normal metal - high $T_{C}$ cuprate junctions 
in the presence of magnetic field}
\def\runauthor{%
Y. \textsc{Tanaka},
H. \textsc{Itoh},
H. \textsc{Tsuchiura},
J. \textsc{Inoue},
Y. \textsc{Tanuma},
and S. \textsc{Kashiwaya}
}
\title{Tunneling conductance
in normal metal - high $T_{C}$ cuprate junctions 
in the presence of magnetic field}
\author{
Yukio \textsc{Tanaka}
\thanks{E-mail address:ytanaka@nuap.nagoya-u.ac.jp},
Hiroyoshi \textsc{Itoh}$^{1}$,
Yasunari \textsc{Tanuma}$^{3}$,
Hiroki \textsc{Tsuchiura}$^{2}$,
Junichiro \textsc{Inoue},
and Satoshi \textsc{Kashiwaya}$^{3}$
}

\inst{%
Department of Applied Physics,
Nagoya University, Nagoya, 464-8603 \\
$^1$Department of Quantum Engineering,
Nagoya University, Nagoya, 464-8603 \\
$^2$Graduate School of Natural Science and Technology,
Okayama University, Okayama 700-8530 \\
$^3$CREST, Japan Science and Technology Corporation (JST),
    Nagoya 464-8603 \\
$^4$National Institute of Advanced Industrial Science
    and Technology, Tsukuba, 305-0045
}

\date{\today}

\abst{%
The magnetic field responses of the zero--bias conductance peak (ZBCP)
in tunneling spectra of high $T_{C}$ cuprate
junctions are studied theoretically.
Our calculation is based on the lattice
Green's function method and takes
the realistic electronic structure of the high $T_{C}$ cuprate
into account.
In marked contrast to previous works,
it is shown that the critical magnetic field strength $H_{C}$
exists for the splittings of the ZBCP's to be discernible.
$H_{C}$ is almost proportional to the
product of the magnitude of pair potential,
transmissivity of the junction,
and the inverse of the Fermi velocity  parallel to
the interface ( 1/$v_{Fy}$ ).
The calculated $H_{C}$'s for
the hole-doped superconductors 
are higher
than those for electron-doped ones
because of relatively large magnitude of pair potential
and the small magnitude of  $v_{Fy}$ originating from
peculiar shape of the Fermi surface.
\par
}

\kword{%
$d$-wave superconductor, Doppler shift,
zero-bias conductance peak, Fermi velocity
}

\begin{document}
\maketitle

\section{Introduction}
Nowadays, almost all of the high-$T_{C}$ cuprates are 
clarified  to be $d$-wave 
superconductors \cite{Sigrist,Harlingen,Tsuei}
and several experiments reported 
the existence of phase coherent phenomena 
peculiar to the $d$-wave symmetry.  
One of this is so called 
the zero bias conductance peak  (ZBCP) 
observed in the tunneling spectroscopy in 
high $T_{C}$ cuprate junctions. 
It was clarified that ZBCP is 
a direct consequence of $d_{x^{2}-y^{2}}$ 
symmetry of the pair potential  
\cite{Kasi00,Hu,Tanaka95,Kashi95,K99,Buchholtz,Barash1,Tanuma1,Wei,Iguchi}.
The presence of ZBCP is a manifestation of the 
formation of the Andreev bound state (ABS)
at the Fermi energy (zero-energy) near a
specularly reflecting  surface \cite{Hu} or interface \cite{Tanaka96}, 
when the angle between the lobe direction of the 
$d_{x^{2}-y^{2}}$-wave pair potential and the  normal to the interface 
is nonzero.
This state originates from the interference effect 
between the injected and reflected quasiparticles
at the  surface or interface
and the sign change of
the $d_{x^{2}-y^{2}}$-wave pair potential. 
There is a large number of studies on ZBCP
\cite{Kasi00,Ekin,Alffm,Biswas,Sawa},
and its related problems
\cite{Covington,A1,A2,Krupke,Dagan1,Dagan2,Matsumoto,Fogel,TJ1,TJ2}.
However, the experimental results of the 
dependence of ZBCP on an external magnetic field 
are not converged yet. 
The splitting of the ZBCP in a magnetic field has been reported 
in some experiments \cite{Covington,Krupke,Dagan1,Dagan2}  
while others can not reproduce it\cite{Ekin,Alffm,Biswas,Sawa}.  
We can consider whether splitting of ZBCP is observed 
or not is quite sensitive to the experimental situations 
and doping concentration of actual high $T_{C}$ cuprates. \par
The effect of screening currents which is so called 
Doppler shift of the energy levels of quasiparticles 
is considered  to be the most promising idea to understand the 
origin of the peak splitting \cite{Fogel,Covington,Krupke}. 
It was pointed out that in an applied magnetic field $H$, 
screening currents shift the ABS spectrum 
\cite{Barash2,Higashitani,Lofwander}
and lead to a 
splitting of ZBCP that is linear in $H$ at low fields \cite{Fogel}. 
The previous theory \cite{Fogel} predicts the splitting of 
ZBCP in any case and is  consistent with 
some experimental results \cite{Covington,A1,A2}.  
However, the absence of ZBCP splitting has not been clarified yet. 
We must need  further study which  takes into account  the 
electronic structures peculiar to actual junctions of 
high $T_{C}$ cuprates.
Recently, four of the aurhors study about 
the effect of transmissivity of the junction on the 
splitting of ZBCP \cite{Tanaka2002}.
It was found there is a critical value of the magnetic field  
$H_{C}$ above which ZBCP splits by external magnetic field 
below which it does not \cite{Tanaka2002}. 
The critical field is almost proportional to the 
width of ZBCP in the absence of magnetic field and 
equivalently to the transparency of the junctions. 
However, in this paper, the shape of Fermi surface 
is assumed to be cylindrical and the reduction of the magnitude of the 
Fermi velocity parallel to the interface 
in the actual Fermi surface is not taken into account. 
Since the degree of the Doppler shift is determined by the 
Fermi velocity parallel to the interface, 
it is natural to expect that the  
shape of Fermi surface influence significantly on the ZBCP splitting.  
In order to understand the actual line shape of the 
tunneling conductance,  
it is indispensable to 
calculate tunneling conductance taking into account 
 the effect of the shape of  Fermi surface.  \par
The aim of this paper is to 
predict the tunneling conductance 
of normal metal - high $T_{C}$ superconductor junctions 
in the presence of magnetic field taking into account 
the shape of  Fermi surface. 
In the present paper,  
based on the Kubo formula 
\cite{kubo,takane92,lee,asano,itoh01},  
we will clarify that 
there is a critical value of the magnetic field  $H_{C}$, 
which crucially determines the peak structure of the tunneling conductance. 
For $H >H_{C}$, ZBCP  splits into two by the magnetic field 
\cite{Fogel}. 
On the other hand, for $H<H_{C}$, ZBCP does not split at least explicitly 
and only the height of ZBCP 
is reduced. 
It is revealed that  
$H_{C}$ is proportional to the product of 
the width $\Gamma$ of ZBCP for $H=0$ 
and the inverse of the magnitude of Fermi velocity in 
$d$-wave superconductor parallel to the interface ($v_{Fy}$). 
Since the magnitude of $\Gamma$ is roughly proportional to  the 
product of the transmissivity of the junctions and the magnitude of the 
pair potential,  $H_{C}$ is enhanced (suppressed) for 
high (low) transmissive junctions and large (small) magnitude of 
the pair potential of $d$-wave superconductors. 
We  also clarify the 
asymmetric behavior of  
$H_{C}$ with respect to the carrier type of the 
high $T_{C}$ cuprates mainly due to the difference in the 
shapes of the Fermi surface. \par
\begin{figure}[htb]
\begin{center}
\scalebox{0.50}{
\includegraphics{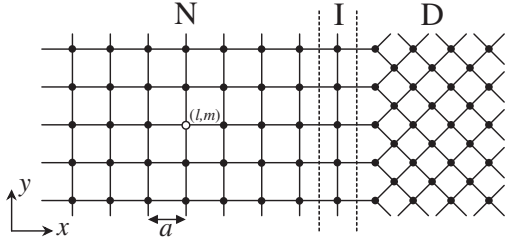}}
\caption{
Schematic illustration of the geometry of the 
normal metal (N)/ insulator (I)/ $d_{x^{2}-y^{2}}$-wave 
superconductor (D) junction. 
N and I are stacked along (100) axis of a square lattice 
while D is stacked along (110) axis. 
Lattice spacing is $a$ for N and I and $a'(=a/\sqrt{2})$ for D. 
\label{fig01}
}
\end{center}
\end{figure}
\section{Formulation}
We consider a $N/I/D$ junction on 2-dimensional lattice system, 
where $N$ and $I$ are stacked along (100) axis of square lattice 
with lattice spacing $a$ 
and $I$ is connected to the (110) surface of $D$ (see Fig.~\ref{fig01}). 
In order to describe the $d$-wave superconductor, 
we use the $t$-$J$ model which is one of the promising models 
for high $T_C$ cuprates. 
By using a mean-field approximation based on the Gutzwiller approximation, 
the model can be mapped into a BCS like mean-field Hamiltonian
\cite{Tsuchiura} $H + \Delta$ with  
\begin{equation}
H = 
 -t_1 \sum_{(\mbox{\boldmath i},\mbox{\boldmath j}),\sigma}
      c_{\mbox{\boldmath i},\sigma}^\dagger 
      c_{\mbox{\boldmath j},\sigma}
 -t_2 \sum_{(\mbox{\boldmath i},\mbox{\boldmath j})',\sigma} 
      c_{\mbox{\boldmath i},\sigma}^\dagger 
      c_{\mbox
{\boldmath j},\sigma} 
- \mu \hat{n} \,,
\end{equation}
\begin{equation}
\Delta = 
\sum_{(\mbox{\boldmath i},\mbox{\boldmath j}),\sigma} 
(\Delta_{\mbox{\boldmath i},\mbox{\boldmath j}}^{\dagger}  
c_{\mbox{\boldmath j},-\sigma}^{\dagger} 
c_{\mbox{\boldmath i},\sigma}^{\dagger} + H.c.), 
\end{equation}
where summations over $(\cdots)$ and $(\cdots)'$ run 
nearest-neighbor and next-nearest-neighbor pairs, respectively. 
The hopping parameters $t_1$ and $t_2$ include so-called 
Gutzwiller factor 
and the pair potential 
$\Delta_{\mbox{\boldmath i},\mbox{\boldmath j}}$ 
has $d_{x^2-y^2}$ symmetry. 
In the middle of the superconductor, 
$\Delta_{\mbox{\boldmath i},\mbox{\boldmath j}}$ is 
$\Delta_{0}(\delta)$ for 
$\mbox{\boldmath j} = \mbox{\boldmath i} \pm (1/2,1/2)$, 
$-\Delta_{0}(\delta)$ for 
$\mbox{\boldmath j} = \mbox{\boldmath i} \pm (1/2,-1/2)$, 
and zero for others, 
where $\delta$ denotes the doping concentration 
of hole or electron. 
Due to the translational invariance along the $y$ direction,  
the creation and annihilation operator 
$c_{\mbox{\boldmath j},\sigma}^{\dagger}$ and 
$c_{\mbox{\boldmath j},\sigma}$ can be expressed as 
$c_{\mbox{\boldmath j},\sigma} 
= \sum_{k_{y}} c_{j_{x},\sigma}(k_{y}) \exp(-ik_y j_{y}a)$. 
Since  the coherence length of the high $T_{C}$ cuprates $\xi$ is 
much  smaller than the penetration depth of the magnetic field 
$\lambda$, and the  effect of the magnetic field on 
the normal metal is not important, 
we can choose the spatial dependence of 
vector potential as  $A_{y}(x)=-H \lambda$, where $H$ is the applied 
magnetic field \cite{Fogel}. 
The quantity $k_{y}$ is replaced with $k_{y}-eH \lambda/\hbar$.  
In the following, the applied 
magnetic field 
is normalized by $H_{0} = \phi_{0}/(\pi^2 \xi_{0} \lambda)$
with $\phi_{0} = h/(2e)$ and 
$\xi_{0}=\hbar v_{F0}/(\Delta_{00} \pi)$. 
The order of the magnitude of 
$H_{0}$ becomes about 1 Tesla when we choose $\xi_{0}$ and 
$\lambda$ as 10\AA and 1500\AA, respectively, 
for hole-doped cuprates with $\delta=0.1$, 
where $\Delta_{00}=0.12t_{0}$ and $v_{F0}=0.16t_{0} a'/\hbar$ 
with the lattice constant $a'$ of cuprates and the unit $t_{0}$ of energy. 

As for the normal region ($N$ and $I$), 
we use the single-orbital tight-binding model. 
The Hamiltonian is 
\begin{eqnarray}
H &=&
  -t \sum_{l,m,\sigma} \left( 
     c_{l+1,m,\sigma}^\dagger c_{l,m,\sigma}
   + c_{l,m+1,\sigma}^\dagger c_{l,m,\sigma}
   + H.c.
  \right) \nonumber \\
  &+& \sum_{l,m,\sigma} 
    v_{l,m} c_{l,m,\sigma}^\dagger c_{l,m,\sigma},
\end{eqnarray}
where $t$ is the hopping integral between nearest neighbor sites and 
$v_{l,m}$ is the on-site potential at a site $(l,m)$. 
Since the system considered has the translational invariance 
in $y$-direction, the conductance $\Gamma_{S}(eV)$ of the junction 
is given by Kubo formula \cite{kubo,takane92,lee}
using the Green's function $G(E)$ as 
\begin{eqnarray}
\Gamma_{S}(E) &=& 2 \frac{e^2}{h} \frac{t^2}{2} 
\sum_{k_y} \left[
  \bar{G}_{l,l+1}^{\,\, k_y} \bar{G}_{l,l+1}^{\,\, k_y}
+ \bar{G}_{l+1,l}^{\,\, k_y} \bar{G}_{l+1,l}^{\,\, k_y} 
  \right. \nonumber \\
&-& \left.
  \bar{G}_{l,l}^{\,\, k_y} \bar{G}_{l+1,l+1}^{\,\, k_y}
- \bar{G}_{l+1,l+1}^{\,\, k_y} \bar{G}_{l,l}^{\,\, k_y}
\right]_{11}
\label{eqn:kubo4} \,, 
\end{eqnarray}
\begin{equation}
\bar{G} = G(E-i0) - G(E+i0) \,,
\end{equation}
\begin{equation}
G(z) = \left(
   \begin{array}{cc}
   z{\bf 1} - H  & \Delta \\
   \Delta^\dagger    & z{\bf 1} + H^*
   \end{array}
\right)^{-1} \,,
\end{equation}
where the relation 
$c_{l, \sigma}(k_y) = \sum_m c_{l,m, \sigma} \exp(ik_yma)$
was used 
and $\bar{G}_{l,l'}^{\,\, k_y}$ is a $2 \times 2$ matrix. 
First, we calculate the isolated Green's function of the $d$-wave 
superconductor and normal metal. Then we obtain the combined Green's 
function by solving Dyson's equation iteratively 
and finally we obtain the tunneling 
conductance $\Gamma_{S}(eV)$ \cite{asano,itoh01}. 
We also calculate the corresponding quantity $\Gamma_{N}(eV)$, 
where the $d$-wave superconductor is in the normal state 
and the pair potential is, then, zero. 
Hereafter, we will look at the normalized value 
$\sigma_{T}(eV)=\Gamma_{S}(eV)/\Gamma_{N}(eV)$.  

In the actual calculation, we choose material 
parameters in the $d$-wave superconductor determined by applying 
the Gutzwiller approximations to $t$-$J$ model. 
For both the normal metal and insulator, 
we use $t=10t_{0}$ and $\mu=0$.  As regards $v_{l,m}$, 
we choose $v_{l,m}=20t_{0}$ for normal metal,  
while for insulator 
we choose $v_{l,m}=35t_{0}$ for the junction with 
high transmissivity, 
and $v_{l,m}=65t_{0}$ for junctions with low transmissivity. 
The corresponding transmissivities $\gamma$ are  given by 
$0.4$ and $0.1$, respectively. 
The thickness of the insulator is chosen as one atomic layer. 

\begin{figure}[htb]
\begin{center}\leavevmode
\scalebox{0.50}{
\includegraphics{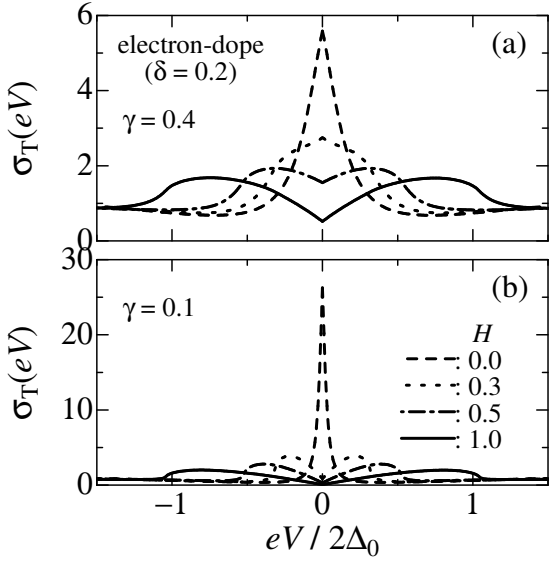}}
\caption{
Tunnel conductance $\sigma_{T}(eV)$ 
of N/ I / D(electron dope, $\delta=0.2$) junctions 
with (a) high ($\gamma=0.4$) and (b) low ($\gamma=0.1$) transmissivities. 
Results obtained for $H = 0.0$, $0.3$, $0.5$, and $1.0$ in a unit of $H_0$ 
are plotted by dashed, dotted, chained, and solid lines, respectively. 
\label{fig02}
}
\end{center}
\end{figure}
\section{Numerical calculation and results}
First, we study the electron over-doped case, 
$i.e.$, $\delta=0.2$, (Fig.~\ref{fig02}), 
where $t_1=0.4150t_{0}$, $t_2=0.1333t_{0}$, 
$\mu=0.0214t_{0}$ for the $d$-wave superconductor\cite{Tsuchiura}. 
The ZBCP splits into two as the magnetic field increases
(see Fig.~\ref{fig02}). 
The degree of the splitting is crucially 
influenced by the transmissivity of the junctions.  
With an increase in transmissivity of the junctions, 
the critical value of magnetic field $H_{C}$ above which 
ZBCP splits into two decreases.

\begin{figure}[htb]
\begin{center}\leavevmode
\scalebox{0.50}{
\includegraphics{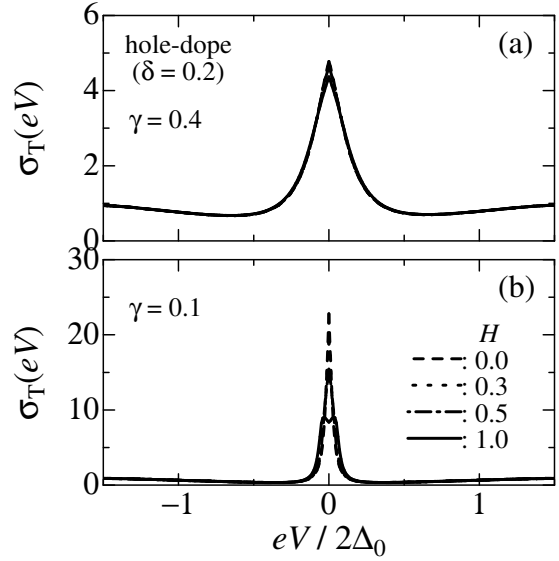}}
\caption{
Tunnel conductance $\sigma_{T}(eV)$ 
of N/ I / D(hole dope, $\delta=0.2$) junctions 
with (a) high ($\gamma=0.4$) and (b) low ($\gamma=0.1$) transmissivities. 
Results obtained for $H = 0.0$, $0.3$, $0.5$, and $1.0$ in a unit of $H_0$ 
are plotted by dashed, dotted, chained, and solid lines, respectively. 
\label{fig03}
}
\end{center}
\end{figure}

In Fig.~\ref{fig03},
corresponding plot for hole over-doped case ($\delta=0.2$) 
is shown, where material parameters are chosen as  
$t_1=0.4117t_{0}$, $t_2=-0.1333t_{0}$, and $\mu=-0.4601t_{0}$ \cite{Tsuchiura}.
For hole-doped materials, the degree of ZBCP splitting is 
weakened, 
and  
ZBCP does not split into two by magnetic field 
except for low transmissivity and high magnetic field junctions. 
As compared to Fig.~\ref{fig02},
the magnitude of $H_{C}$ is drastically enhanced. 

\begin{figure}[htb]
\begin{center}\leavevmode
\scalebox{0.50}{
\includegraphics{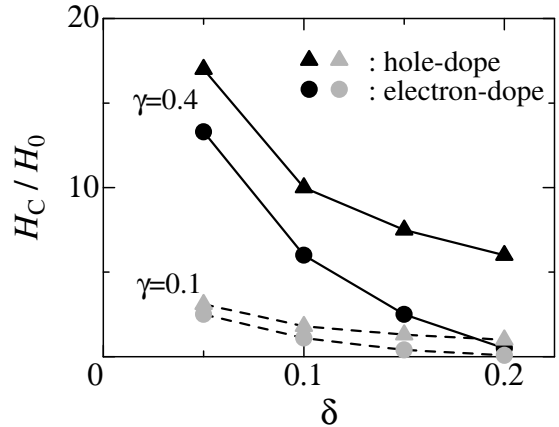}}
\caption{
$H_{C}(\delta)/H_{0}$ calculated for junctions with 
high ($\gamma=0.4$) and low ($\gamma=0.1$) transmissivities 
as a function of the doping rates $\delta$ 
of hole (triangles) and electron (circles). 
Solid and dashed curves are guides for eyes. 
$H_{0}$ is about 1T.  
\label{fig04}}
\end{center}
\end{figure}

To clarify this point,  $\delta$ dependence of 
$H_{C}=H_{C}(\delta)$ 
is plotted in Fig.~\ref{fig04}
both for electron and hole-doped cases 
with low and high transmissivity of the junctions. 
The magnitude of $H_{C}(\delta)$ is a decreasing function 
with $\delta$ both for electron and hole-doped cases. 
In the following, we will consider a microscopic origin of 
$H_{C}(\delta)$.

\begin{figure}[htb]
\begin{center}\leavevmode
\scalebox{0.50}{
\includegraphics{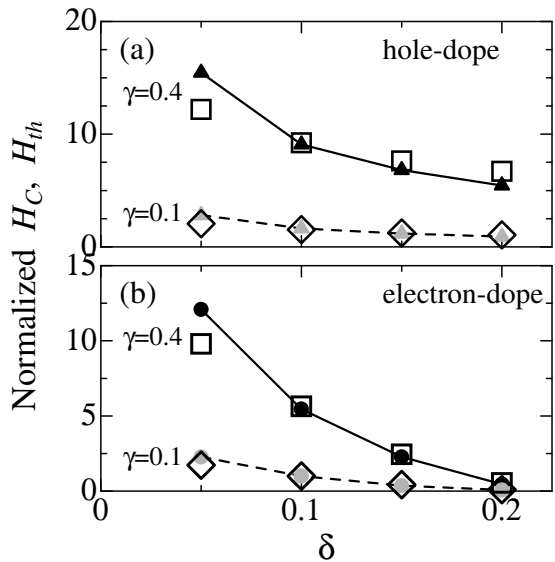}}
\caption{
Normalized $H_{C}$ and $H_{th}$ calculated for junctions with 
high ($\gamma=0.4$) and low ($\gamma=0.1$) transmissivities 
as a function of the doping rates $\delta$ of (a) hole and (b) electron. 
$H_{C}(\delta)/H_{C}(0.1;\mbox{\rm electron-dope})$  and 
$H_{th}(\delta)/H_{th}(0.1;\mbox{\rm electron-dope})$ 
are plotted by solid and open symbols, respectively. 
Solid and dashed curves are guides for eyes. 
\label{fig05}}
\end{center}
\end{figure}

%
In the presence of magnetic field, the energy of the quasiparticle 
$E$ is substituted with 
\begin{equation}
E-ev_{Fy}A_{y}(x)
=E + \frac{H \Delta_{00} v_{Fy}}{H_{0}v_{F0}}. 
\end{equation}
Since positive $H$ is chosen, 
quasiparticles energy $E$ increases (decreases) 
by $H$ for $v_{Fy} >0$ ($v_{Fy}<0$). 
The amplitude of this energy  shift  is 
${H \Delta_{0} < \mid v_{Fy} \mid >}/(H_{0}v_{F0})$, 
where $<\mid v_{Fy} \mid>$ denotes the average of 
the $y$ component of the 
Fermi velocity on the Fermi surface. 
By comparing the energy shift and the 
width $\Gamma$ of ZBCP with 
$\sigma_{T}(\frac{\Gamma}{2})= \frac{1}{2}\sigma_{T}(0)$ 
for $H=0$, 
we can analytically  estimate critical 
value $H_{C}$ which we call $H_{th}(\delta)$ 
to distinguish from $H_{C}(\delta)$ in Fig.~\ref{fig04}. 
The resulting $H_{th}(\delta)$ is 
given by 
\begin{equation}
H_{th}(\delta)=\frac{C \Gamma  H_{0} v_{F0}}{2 \Delta_{00} 
<\mid v_{Fy} \mid> } \,,
\end{equation}
where $C$ is a constant.  
To confirm the correspondence between 
$H_{C}(\delta)$ and $H_{th}(\delta)$, 
we plot $H_{C}(\delta)/H_{C}$ (0.10; electron-dope)
and $H_{th}(\delta)/H_{th}$ (0.10; electron-dope)
in Fig.~\ref{fig05}
for low and high transmissivity of the junctions with 
both electron and hole-doped cases.

\begin{figure}[htb]
\begin{center}\leavevmode
\scalebox{0.50}{
\includegraphics{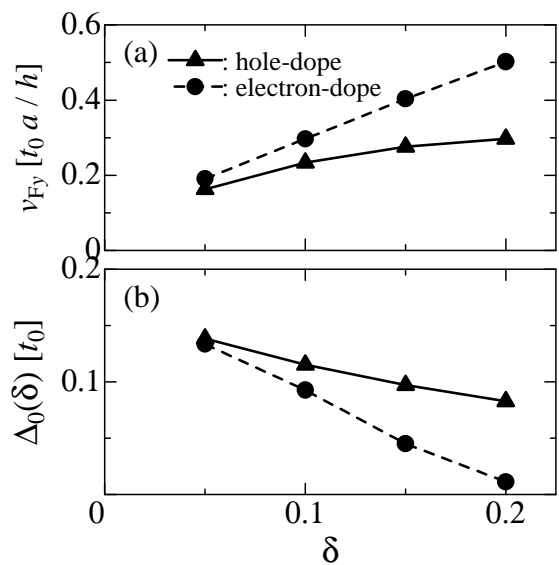}}
\caption{
The magnitude of $v_{Fy}$ and $\Delta_{0}(\delta)$ 
as a function of $\delta$. 
\label{fig06}}
\end{center}
\end{figure}

The coincidence between these two values are fairly well, 
and we can expect  $H_{C}(\delta) \sim H_{th}(\delta)$. 
As discussed above, 
there are two crucial factors which determine the magnitude of $H_{C}$. 
First one is the magnitude of 
$\Gamma$ which 
is almost proportional to the product of the 
transmissivity of the junctions and the magnitude of the 
pair potential $\Delta_{0}(\delta)$.  
The magnitude of $\Delta_{0}(\delta)$ 
is enhanced with the decrease of 
$\delta$ as seen from Fig. \ref{fig06}. 
The second one is the magnitude of 
$< \mid v_{Fy} \mid>$, which is reduced significantly for 
hole-doped high $T_{C}$ cuprates (see Fig. \ref{fig06}). 
In the light of our calculation, the absence of ZBCP splitting 
in the tunneling experiment of hole-doped high $T_{C}$ cuprates 
is fully plausible, since the  magnitude of $H_{C}$ is enhanced. \par

\section{Conclusion}
In this paper, influence of the magnetic field on the 
tunneling conductance in $N/I/D$ junctions  is studied in 
detail where the realistic Fermi surfaces of high $T_{C}$ 
cuprates are taken into account. 
We  concentrated on the splitting of the ZBCP due to 
the magnetic field for (110)-oriented junctions. 
We can define a critical magnetic field $H_{C}$, 
where only 
for $H >H_{C}$, ZBCP splitting occurs by magnetic field \cite{Fogel}. 
It is revealed that $H_{C}$ is proportional to the  
product of the transmissivity of the junction \cite{Tanaka2002}, 
the magnitude of pair potential, and the  
inverse of the  averaged Fermi velocity parallel to the interface.  
In the light of our theory, 
the absence of the splitting of 
ZBCP reported in several experiments \cite{Ekin,Alffm,Biswas,Sawa}
is due to the larger magnitude of $H_{C}$  
originating from the low doping concentration 
or high transmissivity of the junctions. Systematic studies of 
tunneling experiments 
changing doping concentration 
in the presence of magnetic field are desired. \par
In the present paper, the effect of the impurity scattering or surface 
roughness is not taken into account \cite{Tanuma1}. 
These effects broaden the ZBCP \cite{Barash2,Poenicke,Tanaka2001}
and the magnitude of $H_{C}(\delta)$ 
is expected to be enhanced. It is actually an interesting problem to 
clarify the influence of the impurity scattering on 
$H_{C}(\delta)$. \par
In the present paper, 
spatial dependence of the pair potential is not taken into account. 
It is revealed that if the symmetry of the pair potential 
is pure $d$-wave, the low bias property of the tunneling conductance is 
not influenced by the spatial modulation of the pair potential 
\cite{Kasi00}. However, in the presence of the 
broken time reversal symmetry state (BTRSS) near the interface,  
where the mixing of
subdominant $s$- or $d_{xy}$-wave component 
as the imaginary part of the pair potential 
to the predominant $d_{x^{2}-y^{2}}$-wave component 
\cite{Matsumoto,SBL95,Tanuma2,Tanuma3,Sharoni,Sharoni1} 
occurs, the ZBCP splits into two 
even without magnetic field. 
It is also one of  interesting problems to clarify the 
magnetic field dependence of the conductance 
in the  presence of the BTRSS \cite{Tanuma2002}. \par
Finally, we must comment about recent experiments by 
Dagan and Deutsher \cite{Dagan1,Dagan2}. 
They have found that the width of the splitting of 
ZBCP is thickness independent 
for films thinner than the penetration depth. 
This implies that the magnetic field splitting is not 
primarily a Doppler shift effect due to the Meissner screening currents. 
We must seek for other possibilities including 
field induced modification of the pair potential 
\cite{Dagan1,Dagan2,Laughlin}. 

This work is supported by
the Core Research for
Evolutional Science and Technology (CREST) of the Japan Science
and Technology Corporation (JST).
J.I. acknowledges support by the NEDO project NAME. 
The computational aspect of this work has been performed at the
facilities of the Supercomputer Center, Institute for Solid State Physics,
University of Tokyo and the Computer Center.


\begin{thebibliography}{99}
\bibitem{Sigrist}
M.~Sigrist and T.M.~Rice:
Rev. Mod. Phys. {\bf 67} (1995) 505.
%
\bibitem{Harlingen}
D.J.~Van~Harlingen:
Rev. Mod. Phys. {\bf 67} (1995) 515.
%
\bibitem{Tsuei}
C.C.~Tsuei and J.R.~Kirtley:
Rev. Mod. Phys. {\bf 72} (2001) 969.
%
\bibitem{Kasi00}
S.~Kashiwaya and Y.~Tanaka:
Rep. Prog. Phys. {\bf 63} (2000) 1641.
%
\bibitem{Hu}
C. R. Hu:
Phys. Rev. Lett. {\bf 72} (1994) 1526.
%
\bibitem{Tanaka95}
Y.~Tanaka and S.~Kashiwaya:
Phys. Rev. Lett. {\bf 74} (1995) 3451.
%
\bibitem{Kashi95} 
S.~Kashiwaya, Y.~Tanaka, M.~Koyanagi,
H.~Takashima and K.~Kajimura:
Phys. Rev. B {\bf 51} (1995) 1350.
%
\bibitem{K99}
S. Kashiwaya, Y. Tanaka, N. Yoshida and M. R. Beasley:
Phys. Rev. B {\bf 60} (1999) 3572.
%
\bibitem{Buchholtz}
J. Buchholtz, M. Palumbo, D. Rainer and J. A. Sauls:
J. Low. Temp. Phys. {\bf 101} (1997) 1079.

\bibitem{Barash1}
Y. S. Barash, A. A. Svidzinsky and H. Burkhardt:
Phys. Rev. B {\bf 55} (1997) 15282.

\bibitem{Tanuma1}
Y.~Tanuma, Y.~Tanaka, M.~Yamashiro and S.~Kashiwaya:
Phys. Rev. B. {\bf 57} (1998) 7997.
%
\bibitem{Wei}
J.Y.T.~Wei, N.-C.~Yeh, D.F.~Garrigus and M.~Strasik:
Phys. Rev. Lett. {\bf 81} (1998) 2542.
%
\bibitem{Iguchi}
I.~Iguchi, W.~Wang, M.~Yamazaki, Y.~Tanaka and S.~Kashiwaya:
Phys. Rev. B {\bf 62} (2000) R6131.
%
\bibitem{Tanaka96}
Y.~Tanaka and S.~Kashiwaya:
Phys. Rev. B {\bf 53} (1996) 9371.
%
\bibitem{Ekin}
J.W.~Ekin, Y.~Xu, S.~Mao, T.~Venkatesan, D.W.~Face, M.~Eddy
and S.A.~Wolf:
Phys. Rev. B {\bf 56} (1997) 13746.
%
\bibitem{Alffm}
L.~Alff, A.~Beck, R.~Gross, A.~Marx, S.~Kleefisch, T.~Bauch,
H.~Sato, M.~Naito and G.~Koren:
Phys. Rev. B {\bf 58} (1997) 11197.
%
\bibitem{Biswas}
A. Biswas, P. Fournier, M. M. Qazilbash, V. N. Smolyaninova,
H. Balci and R. L. Greene:
cond-mat/0111544.
%
\bibitem{Sawa}
A. Sawa, S. Kashiwaya, H. Kashiwaya, H. Obara,
H. Yamazaki, M. Koyanagi,
I. Kurosawa and Y. Tanaka:
Physica C {\bf 357-360} (2001) 294.
%
\bibitem{Covington}
M.~Covington, M.~Aprili, E.~Paraoanu, L.H.~Greene, 
F.~Xu, J.~Zhu and C.A.~Mirkin:
Phys. Rev. Lett. {\bf 79} (1997) 277.
%
\bibitem{A1}
M. Aprili, M. Covington, E. Paraoanu, B. Niedermeier
and L. H. Greene:
Phys. Rev. B {\bf 57} (1998) 8139. 
\bibitem{A2}
M. Aprili, E. Badica and L. H. Greene:
Phys. Rev. Lett. {\bf 83} 
(1999) 4630. 
%
\bibitem{Krupke}
R.~Krupke and G.~Deutscher:
Phys. Rev. Lett. {\bf 83} (2000) 4634.
%
\bibitem{Dagan1}
Y.~Dagan and G.~Deutscher:
Phys. Rev. Lett. {\bf 87} (2001) 177004.
%
\bibitem{Dagan2}
Y.~Dagan and G.~Deutscher:
Phys. Rev. B {\bf 64} (2001) 092509.
%
\bibitem{Matsumoto}
M.~Matsumoto and H.~Shiba:
J. Phys. Soc. Jpn. {\bf 64} (1995) 3384;
{\it ibid.} {\bf 64} (1995) 4867.
%
\bibitem{Fogel}
M.~Fogelstr\"{o}m, D.~Rainer and J.A.~Sauls:
Phys. Rev. Lett. {\bf 79} (1997) 281.
\bibitem{TJ1}
Y. Tanaka and S. Kashiwaya, 
Phys. Rev. B  {\bf 53} (1996) 11957;  
{\bf 56} (1997) 892; {\bf 58} (1998) 2948. 

\bibitem{TJ2}
Y. Tanaka and S. Kashiwaya, 
J. Phys. Soc. Jpn. {\bf 68} (1999) 3485; 
{\bf 69} (2000) 1152.  

\bibitem{Barash2}
Y.S.~Barash,  M. S. Kalenkov and J. Kurkij\"{a}rvi:
Phys. Rev. B {\bf 62} (2000) 6665.
%
\bibitem{Higashitani}
S. Higashitani:
J. Phys. Soc. Jpn. {\bf 66} (1997) 2556.
%
\bibitem{Lofwander}
T. L\"{o}fwander, V. S. Shumeiko and G. Wendin:
Phys. Rev. B {\bf 62} (2000) R14653.
%
%
\bibitem{Tanaka2002}
Y. Tanaka, H. Tsuchiura, Y. Tanuma
and S. Kashiwaya:
J. Phys. Soc. Jpn. {\bf 71} (2002) 271.
%
%
\bibitem{kubo}
R.~Kubo:
J. Phys. Soc. Jpn. {\bf 12} (1957) 570.
%
\bibitem{takane92}
Y.~Takane and H.~Ebisawa,:
J. Phys. Soc. Jpn. {\bf 61} (1992) 1685.
%
\bibitem{lee}
P.A.~Lee and D.S.~Fisher,
Phys. Rev. Lett. {\bf 47} (1981) 882.
%
\bibitem{asano}
Y.~Asano:
Phys. Rev. B {\bf 63} (2001) 052512.
%
\bibitem{itoh01}
H.~Itoh, Y.~Tanaka, H.~Tsuchiura, J.~Inoue and S.~Kashiwaya:
Physica C, {\bf 367} (2002) 99.


%
%

\bibitem{Tsuchiura}
H.~Tsuchiura, Y.~Tanaka, M.~Ogata and S.~Kashiwaya:
Phys. Rev. Lett. {\bf 84} (2000) 3165. 

%
\bibitem{SBL95}
M.~Sigrist, D.B.~Bailey and R.B.~Laughlin:
Phys. Rev. Lett. {\bf 74} (1995) 3249.
%
\bibitem{Tanuma2}
Y.~Tanuma, Y.~Tanaka, M.~Ogata and S.~Kashiwaya:
Phys. Rev. B {\bf 60} (1999) 9817.
%
\bibitem{Tanuma3}
Y.~Tanuma, Y.~Tanaka and S.~Kashiwaya:
Phys. Rev. B 64, (2001) 214519.

\bibitem{Poenicke}
A. Poenicke, Yu. S. Barash, C. Bruder and V. Istyukov:
Phys. Rev. B {\bf 59} (1999) 7102.

\bibitem{Tanaka2001}
Y. Tanaka, Y. Tanuma and S. Kashiwaya:
Phys. Rev. B {\bf 64}, (2001) 054510.



\bibitem{Sharoni}
A. Sharoni, O. Millo, A. Kohen, Y. Dagan,
R. Beck, G. Deutscher and G. Koren:
cond-mat/0111156.

\bibitem{Sharoni1}
A. Sharoni, G. Koren and O. Millo:
Euprophys. Lett. {\bf 54} (2001) 675.
%

\bibitem{Tanuma2002}
Y. Tanuma, Y. Tanaka and S. Kashiwaya: 
Physica C  {\bf 367} (2002)  147.  

\bibitem{Laughlin}
R.B. Laughlin:
Phys. Rev. Lett. {\bf 80} (1998) 5188.

\end{thebibliography}
\end{document}